%Paper: hep-th/9412145
%From: Martin.Lavelle@IFAE.ES
%Date: Fri, 16 Dec 1994 15:43:19 GMT+0100

  %First some fonts

\font\bigbold=cmbx12
\font\eightrm=cmr8
\font\sixrm=cmr6
\font\fiverm=cmr5
\font\eightbf=cmbx8
\font\sixbf=cmbx6
\font\fivebf=cmbx5
\font\eighti=cmmi8  \skewchar\eighti='177
\font\sixi=cmmi6    \skewchar\sixi='177
\font\fivei=cmmi5
\font\eightsy=cmsy8 \skewchar\eightsy='60
\font\sixsy=cmsy6   \skewchar\sixsy='60
\font\fivesy=cmsy5
\font\eightit=cmti8
\font\eightsl=cmsl8
\font\eighttt=cmtt8
\font\tenfrak=eufm10
\font\sevenfrak=eufm7
\font\fivefrak=eufm5
\font\tenbb=msbm10
\font\sevenbb=msbm7
\font\fivebb=msbm5
\font\tensmc=cmcsc10

%Some Families

\newfam\bbfam
\textfont\bbfam=\tenbb
\scriptfont\bbfam=\sevenbb
\scriptscriptfont\bbfam=\fivebb
\def\Bbb{\fam\bbfam}

\newfam\frakfam
\textfont\frakfam=\tenfrak
\scriptfont\frakfam=\sevenfrak
\scriptscriptfont\frakfam=\fivefrak

%Definition of 8 point

\def\eightpoint{%
\textfont0=\eightrm   \scriptfont0=\sixrm
\scriptscriptfont0=\fiverm  \def\rm{\fam0\eightrm}%
\textfont1=\eighti   \scriptfont1=\sixi
\scriptscriptfont1=\fivei  \def\oldstyle{\fam1\eighti}%
\textfont2=\eightsy   \scriptfont2=\sixsy
\scriptscriptfont2=\fivesy
\textfont\itfam=\eightit  \def\it{\fam\itfam\eightit}%
\textfont\slfam=\eightsl  \def\sl{\fam\slfam\eightsl}%
\textfont\ttfam=\eighttt  \def\tt{\fam\ttfam\eighttt}%
\textfont\bffam=\eightbf   \scriptfont\bffam=\sixbf
\scriptscriptfont\bffam=\fivebf  \def\bf{\fam\bffam\eightbf}%
\abovedisplayskip=9pt plus 2pt minus 6pt
\belowdisplayskip=\abovedisplayskip
\abovedisplayshortskip=0pt plus 2pt
\belowdisplayshortskip=5pt plus2pt minus 3pt
\smallskipamount=2pt plus 1pt minus 1pt
\medskipamount=4pt plus 2pt minus 2pt
\bigskipamount=9pt plus4pt minus 4pt
\setbox\strutbox=\hbox{\vrule height 7pt depth 2pt width 0pt}%
\normalbaselineskip=9pt \normalbaselines
\rm}

%More general stuff

\def\pagewidth#1{\hsize= #1}
\def\pageheight#1{\vsize= #1}
\def\hcorrection#1{\advance\hoffset by #1}
\def\vcorrection#1{\advance\voffset by #1}

\newcount\notenumber  \notenumber=1              %Numbering does
\newif\iftitlepage   \titlepagetrue              %not start on title
\newtoks\titlepagefoot     \titlepagefoot={\hfil}%page
\newtoks\otherpagesfoot    \otherpagesfoot={\hfil\tenrm\folio\hfil}
\footline={\iftitlepage\the\titlepagefoot\global\titlepagefalse
           \else\the\otherpagesfoot\fi}

\def\abstract#1{{\parindent=30pt\narrower\noindent\eightpoint\openup
2pt #1\par}}
\def\smc{\tensmc}

%A nicer footnote

\def\note#1{\unskip\footnote{$^{\the\notenumber}$}
{\eightpoint\openup 1pt
#1}\global\advance\notenumber by 1}

\def\frac#1#2{{#1\over#2}}

\def\tfrac#1#2{{\textstyle{#1\over#2}}}
\def\({\left(}
\def\){\right)}
\def\<{\langle}
\def\>{\rangle}
   %Partial derivatives
\def\2pd#1#2#3{\frac{\partial^2#1}{\partial#2\partial#3}}

\def\sqr#1#2{{\vcenter{\vbox{\hrule height.#2pt
        \hbox{\vrule width.#2pt height#1pt \kern#1pt
           \vrule width.#2pt}
        \hrule height.#2pt}}}}

\def\ni{\noindent}
\def\lqq{\lq\lq}
\def\rqq{\rq\rq}

%%% Macro to generate the equation #'s automatically.
%%% To use start each new section (eg 3) with the commands
%%% \secno=3 \meqno=1 :this will start the equations with (3.1)
%%% Then in place of \eqno(3.1) type \eqn\descriptivename . To refer
%%% back to the equation simply type (\descritivename)
%%% For the appendix set \secno=0, \appno=1\meqno=1 etc
%%% If there are no sections, then set \secno=0

\global\newcount\secno \global\secno=0
\global\newcount\meqno \global\meqno=1
\global\newcount\appno \global\appno=0
\newwrite\eqmac
\def\romappno{\ifcase\appno\or A\or B\or C\or D\or E\or F\or G\or H
\or I\or J\or K\or L\or M\or N\or O\or P\or Q\or R\or S\or T\or U\or
V\or W\or X\or Y\or Z\fi}
\def\eqn#1{
        \ifnum\secno>0
            \eqno(\the\secno.\the\meqno)\xdef#1{\the\secno.\the\meqno}
          \else\ifnum\appno>0
            \eqno({\rm\romappno}.\the\meqno)\xdef#1{{\rm\romappno}.\the\meqno}
          \else
            \eqno(\the\meqno)\xdef#1{\the\meqno}
          \fi
        \fi
\global\advance\meqno by1 }

%%% Macro to assist in the references
%%% At the begining of the paper list the references in the order
%%% that they appear by the command \refn
%%% So if the first reference is to be
%%%  D. McMullan and I. Tsutsui Nucl. Phys. B121 (1994) 12
%%% then type \refn\us{D. McMullan and I. Tsutsui\np{121}{94}{12}}
%%% In the text this is simply refered to by [\us].
%%% At the end of the text type \listrefs

\global\newcount\refno
\global\refno=1 \newwrite\reffile
\newwrite\refmac
\newlinechar=`\^^J
\def\ref#1#2{\the\refno\nref#1{#2}}
\def\nref#1#2{\xdef#1{\the\refno}
\ifnum\refno=1\immediate\openout\reffile=refs.tmp\fi
\immediate\write\reffile{
     \noexpand\item{[\noexpand#1]\ }#2\noexpand\nobreak.}
     \immediate\write\refmac{\def\noexpand#1{\the\refno}}
   \global\advance\refno by1}
\def\semi{;\hfil\noexpand\break ^^J}
\def\nl{\hfil\noexpand\break ^^J}
\def\refn#1#2{\nref#1{#2}}

\def\up#1{$^{[#1]}$}

\def\cmp#1#2#3{{\it Commun. Math. Phys.} {\bf {#1}} (19{#2}) #3}

\def\ijmp#1#2#3{{\it Int. J. Mod. Phys.} {\bf A{#1}} (19{#2}) #3}
\def\mplA#1#2#3{{\it Mod. Phys. Lett.} {\bf A{#1}} (19{#2}) #3}
\def\pl#1#2#3{{\it Phys. Lett.} {\bf {#1}B} (19{#2}) #3}
\def\np#1#2#3{{\it Nucl. Phys.} {\bf B{#1}} (19{#2}) #3}

\def\prl#1#2#3{{\it Phys. Rev. Lett.} {\bf #1} (19{#2}) #3}

%% Some Macros specific to this note%%

\def\d{\delta}

\def\phys{{\hbox{\sevenrm phys}}}

\def\L{{\cal L}}

\def\R{{\Bbb R}}

\def\psiphys{\psi_\phys}

\def\pa{\partial}
\def\vp{\varphi}
%%% Some parameters for this note %%%

\pageheight{24cm}
\pagewidth{15.5cm}
%\hcorrection{-2.5mm}
\magnification \magstep1
\voffset=8truemm
\baselineskip=16pt
\parskip=5pt plus 1pt minus 1pt

%%% The equations

\secno=0

%%% The references
{\eightpoint

\refn\NEWSYMM{M. Lavelle and D. McMullan, \prl{71}{93}{3758}}
\refn\PROPS{M. Lavelle and D. McMullan, \pl{312}{93}{211}}
\refn\CONFINE{M. Lavelle and D. McMullan, \pl{329}{94}{68}}
\refn\DIRAC{P.A.M. Dirac, \lqq Principles of Quantum Mechanics\rqq,
(OUP, Oxford, 1958), page 302}
\refn\SHAB{S.V. Shabanov, \mplA{6}{91}{909}; \pl{255}{91}{398};\nl
 L.V. Prokhorov and S.V. Shabanov, \ijmp{7}{92}{7815}}
\refn\GRIBOV{V.N. Gribov, \np{139}{78}{1};\hfil\nl
I. Singer, \cmp{60}{78}{7}}
\refn\JACK{see, for example, R. Jackiw, in \lqq Current Algebra
and Anomalies\rqq,\nl by
S.B. Treiman et al. (World Scientific, Singapore, 1987)}
\refn\RYDER{see, for example, L.H. Ryder, \lq\lq Quantum Field
Theory\rq\rq, (CUP, Cambridge, 1985)}
\refn\CHO{Y. Choquet-Bruhat, C. Dewitt-Morette and M. Dillard-Bleick,
\lq\lq Analysis, Manifolds and Physics, Revised Edition\rq\rq, (North
Holland, Amsterdam, 1982)}

}
%
%the beginning
%
\rightline {UAB-FT-356}
\rightline {PLY-MS-94-10}
\vskip 40pt
\centerline{\bigbold Observables and Gauge Fixing in Spontaneously
Broken}
\centerline{\bigbold Gauge Theories}
\vskip 30pt
\centerline{\smc Martin Lavelle{\hbox {$^1$}}
and  David McMullan{\hbox {$^2$}}}
\vskip 15pt
{\baselineskip 12pt \centerline{\null$^1$Grup de F\'\i sica Te\`orica
and IFAE}
\centerline{Edificio Cn}
\centerline{Universitat Aut\'onoma de Barcelona}
\centerline{E-08193 Bellaterra (Barcelona)}
\centerline{Spain}
\centerline{email: lavelle@ifae.es}
\vskip 13pt
\centerline{\null$^{2}$School of Mathematics and Statistics}
\centerline{University of Plymouth}
\centerline{Drake Circus, Plymouth, Devon PL4 8AA}
\centerline{U.K.}
\centerline{email: d.mcmullan@plymouth.ac.uk}}
\vskip 7truemm
\vskip 40pt
{\baselineskip=13pt\parindent=0.58in\narrower\ni{\bf Abstract}\quad
Gauge fixing and the observable fields for both abelian and
non-abelian gauge theories with spontaneous breaking of gauge
symmetry are studied. We explicitly show that it is possible to
globally fix the
gauge in the broken sector and hence construct physical fields even
in the non-abelian theory. We predict that any high temperature
restoration of gauge symmetry will be accompanied by a confining
transition.
\par}

\vfill\eject
\noindent In a recent series of papers\up{\NEWSYMM-\CONFINE} we have
investigated the physical degrees of freedom in abelian and
non-abelian gauge theories and the intimately related question of
gauge fixing. For Quantum Electrodynamics (QED) the physical degrees
of freedom are the two transverse photon polarisations and the
observed electron. This electron is not the Lagrangian fermion, which
is neither gauge invariant nor associated with an electric field. In
fact the physical electron is this fermion accompanied by a non-local
photonic cloud\up{\DIRAC}. Use of these physical degrees of freedom
yields a description of QED that is both gauge invariant and infrared
finite already at the level of the Green's functions.

For Quantum Chromodynamics (QCD), or indeed any non-abelian gauge
theory, we have proven\up{\CONFINE} that it is impossible to
globally construct observables describing the fundamental fields (see
also Ref.\thinspace\SHAB).
The obstruction is the Gribov ambiguity\up{\GRIBOV}.  A direct
consequence of this is that it is impossible to observe quarks and
gluons. Local expressions for physical quarks and gluons can,
however, be developed\up{\CONFINE}. The scale at which these
expressions display a gauge dependence is the confinement scale. The
calculation of this scale offers a new approach to determining the
sizes of hadrons.

A question which naturally arises, and which we will address in this
paper, is how can we reconcile the above obstruction to the
observability of the fundamental fields in any non-abelian gauge
theory with the existence of leptons, and the $W$ and $Z$ bosons?
Our resolution
of this apparent difficulty will be to show that in a spontaneously
broken gauge theory one can use the Higgs matter field to fix the
gauge and so circumvent the Gribov ambiguity. We stress that it is
not sufficient merely to introduce scalar matter to achieve this
result, additionally {\it one must be in the spontaneously broken
sector of the theory}. This has consequences for the high temperature
regime.

In QED the physical fields are
$$
A_i^\phys=\(\d_{ij}-\frac{\pa_i\pa_j}{\nabla^2}\)A^j
   \,,\eqn\Aphys
$$
and
$$
\psiphys =\exp\(ig\frac{\pa_iA_i}{\nabla^2}(x)\)\psi(x)   \,,\qquad
\bar\psiphys=\exp\(-ig\frac{\pa_iA_i}{\nabla^2}(x)\)\bar\psi(x) \,.
\eqn\Psiphy
$$
These are straightforwardly seen to be gauge invariant. In scalar
electrodynamics the Lagrangian is
$$
\L= (\pa_\mu+igA_\mu)\phi\,(\pa^\mu-igA^\mu)\phi^{*}-m^2\phi^{*}\phi-
\tfrac14F_{\mu\nu}F^{\mu\nu} \,.
\eqn\Lagscalar
$$
This is invariant under the gauge transformations
$$
A_\mu\to A_\mu^U:=A_\mu+\frac1{ig}U^{-1}\partial_\mu U\,,\qquad\phi\to
\phi^U:=U^{-1}\phi\,,\eqn\gaugetrans
$$
where $U(x)=e^{-ig\Lambda(x)}$.
The physical, gauge invariant scalar fields are
$$
\phi_\phys(x)=\exp\(ig\frac{\pa_iA_i}{\nabla^2}(x)\)\phi(x) \,,
\qquad\phi^{*}_\phys(x)=\exp\(-ig\frac{\pa_iA_i}{\nabla^2}(x)\)
\phi^{*}(x)\,.
\eqn\Scalarphys
$$
They generate the electric field associated with a static charge.
Note that the physical photons are still given
by $(\Aphys)$ in the scalar theory.  These physical fields may be
obtained from the Lagrangian fields by a gauge transformation
$$
A^\phys_i(x)=A^h_i(x)\,,\qquad \phi_\phys(x)=\phi^h(x)\,,
\qquad\phi^{*}_\phys(x)= {\phi^{*}}^h(x)\,,
\eqn\physwithh
$$
where  $h$ is a field dependent  element  of the gauge group that
must itself behave as
$$
h(x)\to h^U(x)=h(x)U^{-1}(x)\,,
\eqn\hbehaves
$$
under an arbitrary gauge transformation. It is readily shown
that these physical fields are then gauge invariant\up{\CONFINE}.
The exact form of $h$ that yields  the above physical  fields may
be read off from $(\Scalarphys)$ and $(\physwithh)$.

In  the  abelian  Higgs  model  one  adds  a  potential   term  $
\lambda(\phi^{*}\phi)^2$  to the  Lagrangian  $(\Lagscalar)$  and
obtains
$$
\eqalign{
\L=(\pa_\mu\phi_1)^2+(\pa_\mu\phi_2)^2&+2gA_\mu\phi_1\pa^\mu\phi_2-
2gA_\mu\phi_2\pa^\mu\phi_1+g^2A_\mu A^\mu(\phi_1^2+\phi_2^2)\cr
&\null\quad-
\lambda(\phi_1^2+\phi_2^2)^2 -\tfrac14F_{\mu\nu}F^{\mu\nu}
  \,,
}\eqn\LagHigg
$$
where    we   have   also   made    a   change    of   variables,
$\phi=\phi_1+i\phi_2$.    To    obtain    spontaneous    symmetry
breaking\up{\RYDER}  one  assumes  that  $m^2$  is  negative.  It
follows that the Higgs potential now develops a minimum at
$$
\vert\phi\vert=a=\sqrt{\frac{-m^2}{2\lambda}}\,.\eqn\Higgscale
$$
To expand around the minimum we write $\phi=\vp+a$.   In terms of
these  new variables  (with the substitution  $\vp=\vp_1+i\vp_2$)
the Lagrangian reads
$$
\eqalign{
\L=(\pa_\mu\vp_1)^2&+(\pa_\mu\vp_2)^2+2gA_\mu\vp_1\pa^\mu\vp_2-
2gA_\mu\vp_2\pa^\mu\vp_1+g^2A_\mu A^\mu(\vp_1^2+\vp_2^2)\cr
&\null\quad+2gaA_\mu\pa^\mu\vp_2
+2g^2aA_\mu A^\mu\vp_1+g^2a^2A_\mu A^\mu-
\lambda(\vp_1^2+\vp_2^2)^2\cr
&{}\null\qquad-4a\lambda(\vp_1^2+\vp_2^2)\vp_1
-4a^2\lambda\vp_1^2 -\tfrac14F_{\mu\nu}F^{\mu\nu}
  \,.
}\eqn\LagHiggShift
$$
In  terms  of  these  shifted  fields  the  gauge  transformation
$(\gaugetrans)$  becomes  $\vp\to e^{ig\Lambda}(\vp+a)-a$,  which
yields
$$
\eqalign{
\vp_1 & \to (\vp_1+a)\cos(g\Lambda)-\vp_2\sin(g\Lambda)-a  \,,\cr
\vp_2 & \to (\vp_1+a)\sin(g\Lambda)+\vp_2\cos(g\Lambda) \,.
}
\eqn\newtrans
$$
Hence the physical, gauge invariant fields with zero vacuum
expectation value are
$$
\eqalign{
{\vp_1}_\phys & =(\vp_1+a)\cos\(g\frac{\pa_iA_i}{\nabla^2}\)-
\vp_2\sin\(g\frac{\pa_iA_i}{\nabla^2}\)-a
 \,,\cr
{\vp_2}_\phys & =(\vp_1+a)\sin\(g\frac{\pa_iA_i}{\nabla^2}\)+
\vp_2\cos\(g\frac{\pa_iA_i}{\nabla^2}\)   \,.
}
\eqn\Shiftphys
$$
Equations (\Aphys) and (\Shiftphys) provide a parameterisation of
the physical  fields in the abelian Higgs model valid in both the
broken ($a\neq 0$) and unbroken ($a=0$) sectors.
Although  these  fields  make clear the physical  content  of the
theory in the unbroken sector, they do not, however,  provide the
most  natural  parameterisation  of the broken  sector,  as these
expressions  reduce  in the Coulomb  gauge  to the transverse
photon (\Aphys) and the pair of scalar fields $(\vp_1\,,
\vp_2)$, thus the expected  content  of the spontaneously  broken
theory, i.e., a massive gauge boson and a single scalar field, is
not immediately  apparent.   In the broken sector  an alternative
basis is, however, also available  since we can now construct the
field dependent group element $h$ from the
scalar matter fields. Choosing
$$
h= \exp\(-i\tan^{-1}\(\frac{\vp_2}{\vp_1+a}\)\)
\,,
\eqn\matterh
$$
which is straightforwardly demonstrated to fulfill $(\hbehaves)$,
we find that
$$
\eqalign{
A^\phys_i & =A_i+\frac1g\pa_i\tan^{-1}
   \(\frac{\vp_2}{\vp_1+a}\) \,, \cr
{\vp_1}_\phys & = \((\vp_1+a)^2+\vp_2^2\)^\frac12-a\,,
}
\eqn\Betterphys
$$
and
$$
{\vp_2}_\phys  = -\sin\(\tan^{-1}
   \(\frac{\vp_2}{\vp_1+a}\)\) (\vp_1+a) + \cos\(\tan^{-1}
   \(\frac{\vp_2}{\vp_1+a}\)\) \vp_2\equiv0
\,,
\eqn\Vanish
$$
where the last relationship  is a trigonometric  identity.   With
respect to these fields the physical content of the broken sector
of the theory becomes transparent.   In particular working in the
unitary  gauge,  where we set $\vp_2$  to zero, we see that these
fields  just reduce to the three physical  photon components  and
one scalar, physical  Higgs field.  From the gauge invariance  of
the  theory  and  of  the  physical  fields  we  know  that  this
interpretation  of the theory must hold in any gauge, although it
may be, as we saw in Coulomb gauge, obscured.

Our obtaining  this  interpretation  of the abelian  Higgs  model
relied upon our ability to use the unitary  gauge and to make the
gauge transformation $(\matterh)$, which  we now want to
investigate.   To this  end we now  recall  that  a gauge  fixing
condition provides a means for picking out a representative  from
each gauge orbit in the configuration space. Clearly a good gauge
fixing  term  should  only pick out one such representative  from
each orbit.  In both the abelian and non-abelian  Higgs model the
configuration  space  of the  Yang-Mills  and  scalar  fields  is
topologically  trivial,  hence a gauge fixing  condition  will be
good  if it does  not \lq\lq  turn  back\rq\rq\  on itself  as it
slices  through   the  orbits.    The  Faddeev-Popov   functional
determinant  provides  a measure  of this:  for a good choice  of
gauge it does not vanish for any configuration.  In a Hamiltonian
approach\up{\JACK}  to the abelian  theory  this  means  that the
Poisson  bracket  of the gauge condition  and Gauss' law must not
vanish for any allowed field configuration.   If it vanishes  for
some    configuration    of    fields    we    have    a   Gribov
ambiguity\up\GRIBOV. It has been shown by Singer\up{\GRIBOV} that
in non-abelian gauge theories, with some mild assumptions  on the
boundary conditions, all gauge fixings in the vector boson sector
suffer  from  such a Gribov  ambiguity.   Indeed  our account  of
confinement  in QCD is based on the fact that if physical  quarks
could  be defined  then they could  be used  to construct  a good
gauge fixing --- in contradiction  to Singer's result.  Since the
unitary  gauge offers us insight  into the physical  fields  in a
spontaneously broken gauge theory, it is natural to now ask if it
is in fact an allowed  gauge or if it suffers  from a Gribov type
problem.

{}From the Lagrangian  $(\LagHiggShift)$
we obtain  the Gauss  law constraint
$$
G(x)=- \pa_i\pi^i+g\pi_{\vp_1}\vp_2-g\pi_{\vp_2}\vp_1-ag\pi_{\vp_2}
\,,
\eqn\Gauss
$$
where  the $\pi_i,  \pi_{\vp_1},  \pi_{\vp_2}$  are the conjugate
momenta to the $A_i, \vp_1, \vp_2$ fields respectively.   In both
the  broken  and unbroken  sectors  of this  abelian  theory  the
Coulomb gauge, $\partial_i  A_i=0$, is an example of a good gauge
condition since $\{G(x), \partial_i A_i(y)\}=\nabla^2\delta(x-y)$
independently of $a$.  This restates the lack of a Gribov problem
with gauge fixing in this abelian theory.

For the unitary gauge we have
$$
\{G(x),\vp_2(y)\}=g(\vp_1(x)+a)\delta(x-y)\,.\eqn\FPunitary
$$
Now we recall that from finite  energy  considerations\up{\RYDER}
we must  have that the scalar  fields  tend  to zero at infinity.
Hence   $\vp_1\equiv   0$  is  an  allowed   configuration,   but
$\vp_1+a\equiv0$  for non-zero  $a$ is not.  This means  that the
unitary gauge, $\vp_2=0$, is acceptable in the broken sector, but
not  in  the  unbroken   one.   We  will   call  this   a  Gribov
problem\note{More properly we should say that the unbroken theory
has a Gribov problem with gauge fixing in the matter sector:  for
this abelian theory there is no Gribov problem if we gauge fix in
the  gauge  sector.}\  with  the  unitary  gauge  in the unbroken
theory.

The existence of a Gribov ambiguity  for the unitary gauge in the
unbroken  sector  would  seem to imply  that this gauge can there
have  at  best  a  perturbative   validity.   However,  from  the
Lagrangian  $(\LagHiggShift)$  we see that the photon  propagator
does  not exist  for vanishing  $a$.  It is thus  clear  that the
unitary  gauge in the unbroken  sector of the abelian Higgs model
is completely unacceptable.
Similarly  the  field  parameterisation  $(\Betterphys)$  in  the
unbroken  sector is not allowed because  of the $\frac1{\vp_1+a}$
term  which  is singular  for vanishing  $a$,  which  shows  that
$(\matterh)$  is not a well-defined  gauge transformation  if the
gauge symmetry is not broken.

The action of the gauge group on the
Higgs fields, in this model, breaks the linear space into
$U(1)$-orbits that are either circles or the exceptional point $\phi=0$.
Symmetry breaking implies that it is one of the non-trivial orbits
that is the vacuum for the theory. Working in polar coordinates
we write
$\phi=(\vp_1+a)+i\vp_2=\rho\exp i\theta$, and the physical fields
(\Betterphys) are then
$$
\eqalign{
A^\phys_i & \to A_i+\frac1g\pa_i\theta \,, \cr
{\vp_1}_\phys & \to \rho-a \,,
}
\eqn\radials
$$
which simplifies even further in the unitary gauge, $\theta=0$.
The new coordinates, $(\rho, \theta)$, are  ill defined at the origin
$\rho=0$. However, as we have seen, this point is only an allowed
configuration  for the unbroken theory. Hence the polar coordinates
are a globally valid coordinate  system in the broken sector of the
theory, and this is what allows us to find the physical
fields~(\radials).

Before giving the details of how this abelian example can be extended to
the non-abelian theory, it is useful to give a qualitative account of
how
we are to proceed. We know that in an abelian gauge theory we can fix
the
gauge and hence construct physical fields. In a non-abelian theory, the
Higgs  mechanism is used to
reduce the symmetry; so if it can be reduced to
an abelian group (as it is in the electro-weak theory) then we would
expect to be able to construct physical fields.

More geometrically, in  a non-abelian theory with structure group $G$,
the Yang-Mills field is identified with the  connections
on a principal $G$-bundle $P$ over the space time.  As long as the
structure group is non-abelian there will be a Gribov problem associated
with gauge fixing, and hence an obstruction to constructing physical
fields.
The Higgs fields ${\underline\phi}$
takes values in some vector space $V$ and can thus be viewed as cross
sections of the associated bundle $P\times_{_G}V$ ---
with fibres over the
space-time now being the vector space $V$.  The vector space $V$ can be
thought of as a collection of G-orbits, a typical example being when
$G={\rm SO}(n)$, and $V=\R^n$; in which case the orbits are the
origin (with stability group $H=G$) and the $(n-1)$-spheres
(with stability group $H={\rm SO(n-1)}$). In symmetry breaking,
the potential energy of the Higgs
fields is such that the vacuum configuration corresponds to the
Higgs fields
being restricted to one of these orbits, which we identify with the
coset space $G/H$, for some stability subgroup $H$ of $G$. Choosing
a point on an orbit is equivalent to gauge fixing in the matter sector.
These vacuum solutions now correspond to cross sections of the
associated bundle $P\times_{_G}G/H$, which can be identified with the
quotient  $P/H$. Now cross sections of $P/H$ correspond to reductions
of the structure group from $G$ to $H$ (see, for example, the theorem on
page 385 of Ref.\thinspace\CHO). Thus, if we can use gauge fixing in the
matter sector to reduce the structure group to an abelian group, then
the residual gauge symmetry can be dealt with  in the gauge sector
and we will not encounter any Gribov problem. We shall now demonstrate
through an explicit calculation that this is what happens in the
Salam-Weinberg model.

We now consider the case of spontaneously broken SU(2) gauge theory.
The scalar fields $\underline\phi$ are now two-component, complex column
 vectors. The orbits in the target space are three spheres
which we can identify with
SU(2), thus having a trivial stability group.
(This is essentially the weak sector of the standard model with the
Weinberg angle set to zero for simplicity.)  The gauge
transformations of the fields are
$$
\eqalign{
A_i\equiv A_i^a\tau^a\to A^{U}&= UA_iU^{-1}-\frac ig
U\partial_iU^{-1}\,,
\cr
\underline\phi\to\underline\phi^{U}&= U\underline\phi\,.
}
\eqn\nonabtrans
$$
The shifted scalar fields $\underline\vp=\underline\phi-\underline a$
are taken to have zero expectation value, and we explicitly write
$$
\underline\vp=
\pmatrix{
\vp_1+ i \vp_2\cr
\vp_3+ i \vp_4
}- \pmatrix{0\cr a}\,,\eqn\no
$$
where $a$ is real. Under the gauge transformation (\nonabtrans), these
shifted fields transform as
$$
\underline\vp\to\underline\vp^{U}=U(\underline\vp+\underline a)
-\underline a\,.\eqn\no
$$
The physical fields are then identified with
$$
\eqalign{
A_i^\phys=A^h&:=hA_ih^{-1}-\frac{i}{g}h\partial_ih^{-1}\cr
\underline\vp_\phys=\underline\vp^h&:=h(\underline\vp+\underline a)
-\underline a\,,
}\eqn\nonabphy
$$
where the field dependent group element $h$ satisfies (\hbehaves).
A direct calculation shows that
$$
h=\frac1{\sqrt{ \vp_1^2+\vp_2^2+(\vp_3+a)^2+\vp_4^2  }}
\pmatrix{\vp_3+a+i\vp_4 & -\vp_1-i\vp_2\cr
\vp_1-i\vp_2 & \vp_3+a-i\vp_4}
\,,\eqn\nonabh
$$
satisfies this requirement,  and can hence be used to generate
gauge invariant fields. From (\nonabphy)
and (\nonabh) we find that the physical gauge
and scalar fields are
$$
\eqalign{
\underline\vp_\phys  = & \pmatrix{0\cr
(\vp_1^2+\vp_2^2+(\vp_3+a)^2+\vp_4^2)^{\frac12}-a }
\,,\cr
A_{i_\phys}^1=
& \frac1{\Phi^2}\biggl[
\biggr.
-A_i^1 (\vp_1^2-\vp_2^2-\vp_3^2+\vp_4^2)
  +2A_i^2(\vp_1 \vp_2 +
\vp_3\vp_4) +2A_i^3(\vp_1\vp_3-\vp_2\vp_4)\biggl.\biggr] \cr
& +\frac1{g\Phi}
\biggl[
\vp_4\partial_i\vp_1-\vp_1\partial_i\vp_4+
\vp_3\partial_i\vp_2-\vp_2\partial_i\vp_3
\biggr]
\,,\cr
A_{i_\phys}^2
=
& \frac1{\Phi^2}\biggl[
\biggr.
A_i^2 (\vp_1^2-\vp_2^2+\vp_3^2-\vp_4^2)
  +2A_i^1(\vp_1\vp_2 -
\vp_3\vp_4) -2A_i^3(\vp_2\vp_3+\vp_1\vp_4)\biggl.\biggr] \cr
& +\frac1{g\Phi}
\biggl[
\vp_3\partial_i\vp_1-\vp_1\partial_i\vp_3+
\vp_2\partial_i\vp_4-\vp_4\partial_i\vp_2
\biggr]
\,,\cr
A_{i_\phys}^3  = & \frac1{\Phi^2}\biggl[
\biggr.
-A_i^3 (\vp_1^2+\vp_2^2-\vp_3^2-\vp_4^2)
  - 2A_i^1(\vp_1\vp_3 +
\vp_2\vp_4) +2A_i^2(\vp_2\vp_3-\vp_1\vp_4)\biggl.\biggr] \cr
& + \frac1{g\Phi}
\biggl[
\vp_4\partial_i\vp_3-\vp_3\partial_i\vp_4+
\vp_2\partial_i\vp_1-\vp_1\partial_i\vp_2
\biggr]
\,.}
\eqn\nonabelianphys
$$
where $\Phi^2=\vp_1^2+\vp_2^2+(\vp_3+a)^2+\vp_4^2$.
The physical scalar fields are those of the unitary gauge. It is easily
seen that, in the broken sector, this is indeed a good gauge. Recalling
that the Faddeev-Popov matrix is the Poisson bracket of Gauss' law, the
generator of infinitesimal gauge transformations,
with the gauge function, it is easily seen to be invertible. For
example, we have
$$
\{ G^a\epsilon^a,\vp^1\}\sim g\left(
\epsilon_3\vp_2 +\epsilon_1\vp_4- \epsilon_2(\vp_3+a)
\right)
\eqn\nonabFP
$$
and it is clear that the unitary gauge is a \lq good\rq\ gauge even in
this nonabelian model.
In unitary gauge the physical bosons reduce to
$\underline\vp_\phys=\pmatrix{0\cr\vp_3\cr}$ and $A_{i_\phys}^a=
A_i^a$, as one would expect.

Our construction of these physical, gauge-invariant fields in a
non-abelian theory depended crucially upon our ability to fix the
unitary gauge. Since we expect gauge symmetry to be restored at some
finite temperature, it will then no longer be possible to fix a gauge
in  the  scalar sector at all. In the vector boson sector of a
non-abelian theory only a perturbative gauge fixing is
possible\up{\CONFINE} and so
only perturbatively physical fields can be constructed in the
unbroken sector.

In the full Salam-Weinberg  model the fact that we
are dealing  with a non-simple  group means that we are left
with an unbroken  $U(1)$ symmetry  where gauge fixing  is possible
and a complete gauge fixing may be constructed by a trivial extension
of the above analysis:  partially  in the abelian gauge boson sector
and the rest, as above, in the scalar,  Higgs sector.   Thus we
may circumvent  the Gribov ambiguity  in the non-abelian  part of
the Salam-Weinberg  model and build up physical,  massive $W$,
$Z$ and Higgs bosons in the manner of $(\nonabelianphys)$. We
also predict that at high temperatures in the unbroken sector of the
weak interaction  not only the non-abelian vector bosons
and scalars but also all weakly interacting fundamental particles,
such as the electron, will be confined.

\bigskip
In conclusion  we have seen that, when coupling  to scalar  fields,
one can fix the gauge in the scalar  matter  sector  if the gauge
symmetry  is spontaneously  broken.   Since  this holds  even for
non-abelian  gauge theories,  it offers a mechanism  for avoiding
the Gribov ambiguity  in the broken sector of the standard model.
This ability  to fix the gauge allows  us to construct  physical,
gauge invariant fields.  In particular is is possible to describe
the $W$ and $Z$ bosons  so long as the gauge symmetry  is broken.
This is to be contrasted  with the impossibility  of constructing
physical  quarks  and gluons  in QCD.   Our ability  to construct
physical  fields to describe  electrons,  photons and the like in
QED and their equivalents  in the broken  sector  of the standard
model together with the lack of physical  quarks and gluons is in
satisfyingly   complete  agreement  with  experiment.    At  high
temperature  when the gauge symmetry is restored  we predict that
the  $W$  and  $Z$  bosons  and  all  other  fundamental   weakly
interacting  fields  will  not be observables.   This could  have
consequences for studies of the early universe.

\bigskip
\noindent {\bf Acknowledgements}
\smallskip
\noindent ML was partially supported by research project
AEN-93-0474 and thanks the
University of Plymouth
for their hospitality. We thank Sergey Shabanov for correspondence.

{
    \vfill\eject
     \immediate\closeout\reffile%\parindent=20pt
  \centerline{{\bf References}}\bigskip\eightpoint\frenchspacing%
  \input refs.tmp\vfill\eject\nonfrenchspacing}
\bye